\documentclass[10pt, fleqn, twocolumn]{article}
\input epsf.tex
\usepackage{times}
\usepackage{ifthen}
\usepackage[super,numbers,sort&compress]{natbib}
\bibpunct{}{}{,}{s}{}{;}
\bibfont{\small}
\begin{document}

\begin{titlepage}
\title{Gravitational Lens Systems to probe
Extragalactic Magnetic Fields}

\author{D. Narasimha$^1$\thanks{E-mail:
dna@tifr.res.in (DN)}
and S.M. Chitre$^2$\thanks{kumarchitre@hotmail.com (SMC)} \\
 $^1$Tata Institute of Fundamental Research, Mumbai 400 005, India \\
 $^2$Department of Physics, University of Mumbai, Mumbai 400 098, India}

\date{Accepted  .... Received ..... ; in original form ....}

\maketitle

{\parindent=0cm\leftskip=1 cm
\begin{abstract}
The Faraday rotation measurements of multiply-imaged gravitational lens
systems can be effectively used to probe
the existence of  large-scale ordered magnetic fields in lensing galaxies
and galaxy clusters.
The available sample of lens systems appears to suggest the presence of a
coherent large-scale magnetic field in giant elliptical galaxies 
somewhat similar to the spiral galaxies.
\end{abstract}

keywords:
gravitational lensing -- galaxies: magnetic fields -- polarization.

}
\end{titlepage}

\section{Introduction}

The origin of extragalactic magnetic fields is one of the most challenging 
problems in astronomy. Clearly, the detection and measurement of large-scale 
magnetic fields in cosmological objects would be important for our 
understanding of their role in theories of galaxy formation and  
evolution. 
It is hoped that observations of extragalactic objects and high redshift
galaxies can, in principle, illuminate the basic issues relating to their
origin and dynamical amplification.

Magnetic fields in external galaxies have been measured
using radio observations of their synchrotron emission to have strengths of 
about microgauss and they have been detected to have a coherence over a
scale of few kiloparsecs. The large-scale ordered magnetic fields in several
spiral galaxies were reported by\citet{b1}
to have average field strength 
$\sim$ 10 $\mu$G with a coherence scale of several kpc. However, it is
desirable to have completely independent tools to ascertain 
the global characteristics of the pervading magnetic fields in galaxies,
both spiral and elliptical, at a much earlier epoch.
Equally, estimation of the strength of galactic magnetic fields over 
a range of redshifts will be
highly valuable for understanding their origin and evolution with the 
age of the Universe.

 Cosmological magnetic fields could be generated in the early universe by some 
mechanism such as the first-order phase transition \citet{b2},
 coupling of electromagnetic field with curvature \citet{b3}
or by a thermal battery operating in expanding ionization/shock fronts 
impinging on density inhomogeneities in the intergalactic medium 
\citet{b4, b5}.
But these seed field values ($\leq$ 10$^{-19}$G) need to be enhanced 
and maintained at the observed microgauss level by some kind of a dynamo 
process. 
We note parenthetically that the energy density of average galactic magnetic fields is of the 
same order as the energy density of cosmic rays of intergalactic thermal 
energy, and of cosmic microwave background radiation 
(of order 10$^{-13}$ erg/cm$^3$).

The radiation emitted by distant sources, during its passage over 
cosmological distances, is likely to encounter a variety of objects 
en route such as galaxies, galaxy--clusters, Ly$\alpha$ clouds,
magnetic fields and metal line 
absorbers. The imprints left by these intervenors in the form of spectral 
absorption features and Faraday rotation of the polarized flux of the 
background source can, in principle, furnish valuable information about 
the chemical composition or magnetic fields associated with the intervening 
objects. The observed correlation of the Faraday rotation measure (RM) of 
high red-shift quasars with the optically detected absorption-line systems 
along the sightlines prompted
Kronberg and Perry (1982)\citet{b6}
 to estimate the magnetic 
field strength in high redshift objects. For studying  magnetic fields 
in high redshift galaxies and galaxy clusters we should first identify 
extragalactic radio sources with polarized flux that are located
within or behind 
these intervening deflectors and measure the Faraday rotation of radio waves 
coming from the background source. The Faraday rotation will naturally 
have contributions from (i)  our Galaxy, (ii) intervening objects and 
absorption systems, and (iii) the source itself. Clearly, for inferring the 
average strength of high red shift magnetic fields, it is essential to 
subtract out the contributions to the Faraday rotation occurring at the 
source and in our own Galaxy. We need, of course, to have a reasonably 
independent estimate available of the electron column density in the 
intervening objects. The connection between damped Lyman
systems and galaxies  and estimation of the hydrogen column density
from the optical lines
is extensively discussed in the literature (\citet{blasi:99}, \citet{sarg:89})
  As discussed by
\citet{blasi:99} the electron
column density could be assumed to follow the corresponding density of 
neutral hydrogen in the
intervenor which can be estimated from absorption line strengths.

All these requirements may be very conveniently fulfilled for the 
case of radio-selected gravitationally lensed sources. 
In gravitational lens systems we often encounter polarized radio sources 
(e.g. quasars, radio galaxies) that are being multiply imaged by an 
intervening `normal'
galaxy or a galaxy cluster. In such lensed systems the difference in the 
rotation measures between various images is not expected to be severely 
affected by the background source or by our Galaxy, except for possible 
contributions from absorption systems located en route and 
perhaps, contamination from small-scale inhomogeneities in our own  Galaxy. 
In short, because there is  more 
than one sightline to the polarized radio source available for a multiply 
imaged system, it should  be possible to filter out contributions 
from the source and our Galaxy by taking differences between rotation measures 
of various images. We propose to apply this technique for deducing the 
estimates of magnetic fields in galaxy and cluster lenses and particularly
enquire about the nature and strength of magnetic fields in elliptical
galaxies. It will be valuable if we can adopt this technique for
inferring the large-scale magnetic fields in the intergalactic medium
and even more illuminating if we can use the polarized
Cosmic Microwave Background as a source for probing the cosmic
magnetic fields. We can only hope this may conceivably become feasible  with 
rapidly advancing technology.

\section{Inferred Magnetic fields in selected gravitational lens systems}

\begin{figure}
\vskip -0.3 cm
\centerline{\epsfysize=13.5cm\epsfbox{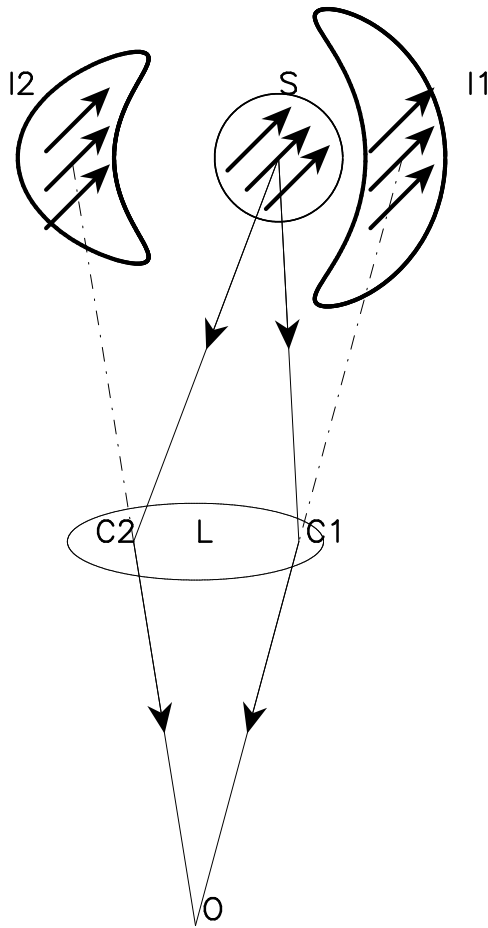}}
\vskip -0.2 cm
\caption{Schematic diagram of the phenomenon of gravitational lensing.
Background source S is lensed by the intervening galaxy L and multiple images
I1 and I2, having identical intrinsic properties are formed.
The regions of the lens sampled by the light rays forming images I1 and I2,
shown by C1 and C2, are separated by typically a few kiloparsecs.}

\end{figure}

For the sake of illustration, we have sketched the phenomenon of
gravitational lensing in Figure 1.
A polarized source such as a quasar, S, 
is lensed by an interveining galaxy L producing the multiple images
I1 and I2.
The phenomenon of gravitational lensing preserves surface brightness and 
also the polarization properties of the original lensed source. 
The fractional polarization as well as the direction
of the electric field in the images should follow the value in the source; 
we have, therefore, shown
the polarization vectors with same length and direction in the source as well
as in all the images. However, the path of the light rays forming the two images
sample different regions of the 
lens galaxy (shown through C1 and C2 in the figure)
and hence will be affected differently by the interstellar medium of the lens.
It was, therefore, recognised after the discovery of the first 
gravitational lens system Q0957+561
\citet{walsh:79} that radio 
observations of such lens systems could furnish valuable information about 
the physical properties of the intervening lens and of absorption system 
along the lines of sight.

 The magneto-ionic plasma in the intervening 
lenses is expected to cause Faraday rotation of the radiation which will, 
of course, vary for each of the light paths from various images. 
The angle of rotation of the plane of polarization is given by
\begin{center}
$$ \Psi_F= {e^3\over 2\pi m_e^2 c^4}\int B_\parallel (l) 
n_e(l) \lambda^2 dl,   \hskip 1.5 cm(1)
$$ 
\end{center}
where  n$_e$ is the electron number density, $\lambda$ is the wavelength
of the radiation as seen by the absorber medium,
 $B_\parallel$  is the line of sight 
component of the magnetic field and the integral is over the path length 
through the intervening absorbers.

The rotation measure, as measured by the
observer at redshift of zero is given by
\begin{center}
$$ RM={\Psi_F\over \lambda^2_{obs}}= {e^3\over 2\pi m_e^2 c^4}\int B_\parallel (l) 
n_e(l)\left[{\lambda(l)\over \lambda (obs)}\right]^2 dl   \hskip 0.5 cm (2)
$$ \end{center}

For the Faraday Rotation produced by a deflector
at redshift $z$, the rotation measure of the
intervening galaxy
with the average line of sight magnetic field component,

\begin{center}
$$<B_\parallel> = {\int n_e (z) B_\parallel (z) dl (z)\over
\int n_e (z) dl (z)} \hskip 2.3 cm (3) $$
\end{center}
and the electron column density, $N_e = \int n_e (z) dl (z)$
 may be expressed as 
\begin{center}
$$ RM  \simeq 2.6\times (N_e)_{19} <B_\parallel>_{\mu G}/(1+z)^2 { }
rad\; m^{-2}. \hskip 0.2 cm (4 ) $$
\end{center}
Here $(N_e)_{19}$ is expressed in units of 10$^{19}$ cm$^{-2}$ and 

\noindent
$<B_\parallel>_{\mu G}$ in units of microgauss. 
Even though the Faraday Rotation may be caused by the source, the intervenor 
and the Milky Way, the difference in the rotation angle between the
multiple images is practically due to the lens which is contained in
Eq. (4). Consequently,
the magnitude of the difference in rotation measures (RM) between images 
turns out to be a valuable probe for estimating the average line of 
sight component of magnetic field in the lenses.

There have been a number of multi-frequency VLA polarization observations of 
gravitational lens systems through the 1980s and 1990s. 
The Faraday rotation measures 
and intrinsic polarization angles of the multiple images of some selected 
lenses
\citet{subh:90, patn:93, patn:99, patn:01, nar:01, king:97, chen:93, p30}
are summarised in column 3 of Table I, 
while column 4 is the best fit estimation of
the differential Faraday Rotation Measure between various images. The last 
column denotes the difference between polarization angles at 
zero wavelength which, in principle, would have the value 0,
were there no Faraday rotation in the source.

\begin{table*}
\centering
\caption {Faraday Rotation in selected Lens systems }
\vspace{1cm}
\begin{tabular}{|c|c|c|c|c|c|c|c|c|c}
\hline
         System   & Lens   &  RM &  Diff. RM 
& Excess &$\chi^2_{**}$&no. of &Time &References&     \\
        &redshift &(rad m$^{-2}$)&(rad m$^{-2}$) & P.A. &&degree&delay && \\ 
        & && & (degree)&&of& && \\ 
        & &(literature)&(best fit)$^*$& ($\lambda$=0)&$ $&freedom&(days)& & \\ \hline
&&&&&&&&&\\
Spirals &&&&&&&&&\\
&&&&&&&&&\\
  B0218+357   &  0.684       & A-8920     & AB: 913$\pm$31& -10&0.3&2&10.5&
\citet{patn:93}& \\
                  &           & B-7920    &  &&& &&&\\ 
&&&&&&&&&\\
 PKS1830--211 & 0.89 & A -157 & AB: 1480$\pm$83&24&7&2&26&
\citet{subh:90}& \\
 && B 456  &&&&&&& \\ \hline
Ellipticals&&&&&&&&&\\
&&&&&&&&&\\
   Q0957+561    &  0.36           & A-61$ \pm$3    & AB: 99$\pm$2& -2&0.3&1&&&   \\
                &          & B -160$\pm$3   & &&&& 417& \citet{green:85}& \\
&&&&&&&&&\\
  B1422+231  &  0.31     & A -4230$\pm$60   &AB: 125$\pm$125& 4.7&16.1&2&  1.5&
\citet{patn:99},& \\
                  &        & B -3440$\pm$88    &AC: 20$\pm$70 & 3.4&5&2&7.6&
\citet{patn:01}& \\
                  &   & C -3340$\pm$90  & BC: 105$\pm$77&-1.3&6.1&2&8.2&& \\ \hline
&&&&&&&&&\\
 1938+666 & 0.878           & A 665$\pm$ 14      & AB: 960$\pm$202 & -26&27&1&&
\citet{king:97}& \\
            &                 & B 465$\pm$ 14     & BC1:85$\pm$39&-10.5&19&1&&&  \\
            &                 & C1 441$\pm$ 3      &C2C1: 56$\pm$4 & -1.1&2.6&3&&& \\ 
            &                 & C2 498$\pm$ 3      & &&&&&& \\ 
 &&&&&&&&&\\
&&&&&&&&&\\
 MG1131+0456 & 0.844 & R1 910  &R1R4: 1200$\pm$ 10&&&&& \citet{chen:93}& \\
 && R2 -72$\pm$ 25&&&& &&& \\
 && R3 18$\pm$13&& &&&&& \\
 && R4 -290$\pm$13 &&&&&&& \\
 && R5 -308$\pm$19 &&&&&&& \\
 && R6 56$\pm$20 &&&&&&& \\
&&&&&&&&&\\ \hline
\end{tabular}

\noindent
$^*$In some cases, the differential RM listed in column 4 are at variance with 
the direct difference because of the ambiguity modulo $\pi$.

\noindent
$_{**}$See the text for discussion on the $\chi^2$.
\end{table*} 

\subsection{Effects of Substructures}

Most of the radio sources have substructures like core, knot and jet
and generally
the source polarization vector among these components is not aligned.
A good illustration of the change in polarization angle across the
substructures can be found \citet{bigg:03}
for the 8.4 GHz VLBA images of the lens system B0218+357
The relative flux 
contribution between these components also happens to change
 gradually with frequency.
There is an added uncertainty introduced in the estimation of Faraday rotation
 from the position angles of the polarization vector  if the measurements
at various frequencies are not made simultaneously.

The change in polarization vector with time, for example as illustrated by
Patnaik \& Narasimha \citet{nar:01}
can also be  a factor.
But, in principle, this uncertainty can be eliminated by getting the maps
at similar spatial resolution and at close epochs and correcting for the
time delay between the images. 
Here we argue that (a) if we ignore the effects of non--simultaneous 
measurements of the position angle,  (b) further make the reasonable 
assumption that the Faraday rotation introduced
by an intervening object does not vary considerably at milliarcsecond scale,
and (c) the source variability and the time--delay together do not
seriously affect the differential Faraday Rotation measurements,
 then we can estimate the difference between the Rotation Measure between the
lines of sight along the multiple images of a background source.

\subsection{Effects of Faraday Rotation in the Milky Way}

The Milky Way is amenable to detailed analysis of its magnetic
field structure due to pulsar and other observations. There is evidence for 
the presence of
magnetic field as well as its direction reversal from very small scale
to the global scale of kiloparsec. However, there is a controversy about
the magnetic field reversal due to difficulties in the analysis
of Faraday rotation measurements specially when the line of sight passes through multiple spirals
\citet{rand:89}, \citet{b39}.
Nevertheless, it might be fair to accept that the Milky Way has
(a)an ordered component of magnetic field of 2 to 10 $\mu$G on the scale of spiral arms,
with the field strength  generally increasing as we go 
inwards to the central regions of the Galaxy and (b) at scales of star forming regions
or stellar environment ($\sim$ 10$^{15}$ cm), there is evidence for
milligauss magnetic field.
Consequently, we could expect differential Rotation Measures of
a few tens of rad m$^{-2}$ if we pass through a star forming region. However,
it could also produce substantial depolarization on subarcsecond scale and
a change in the direction of polarization.
More importantly, in such a case we are unlikely to observe the lensed images.
On the other hand, near the galactic plane we might expect
Faraday Rotation due to the global magnetic field almost aligned along
the spiral arms. But the differential Faraday rotation measure between the
arcsecond scale images of gravitational lens systems would be
marginal at less than about 10 rad m$^{-2}$, 
and could be safely ignored unless the image separation is
tens of arcseconds. Nevertheless, this becomes important, for instance,
when we attempt to estimate magnetic fields of galaxy--clusters
using differential Faraday rotation between multiple images
produced by the gravitational lensing action of the cluster. 
Consequently, it would be difficult to separate effects
 due to our Galaxy and an intervening galaxy-cluster located
almost along the Galactic plane, unless the cluster magnetic field is at least
of the order  of 100 nG.

\section{Discussion}

\subsection{Polarization vector in Lens systems as a Cosmological probe}

\citet{nair:93} 
had, indeed, pointed out suitability of the polarization vector as a 
cosmological probe. The thrust of the analysis was that
modification of the position angle due to inhomogeneities
such as individual stars or gas clouds in the lens would have no role
to play 
provided the scale over which the polarization direction changes
is always larger than the scale over which the inhomogeneities can affect the
background source. The problem will, of course,
be serious if the scale length of the
source polarization vector and the lens inhomogeneity conspire to 
become comparable. 
It is evident from the discussion in the previous section that
the depolarization and Faraday rotation introduced in our own Galaxy
will be similar between the multiple images with separation of the
order of arcseconds. Thus,
if the depolarization is substantial  when an image is
intercepted by a cloud having partially ionized matter,  the
extinction in optical wavelengths would be noticeably high.
The image B in PKS1830--211,
for instance, could be one such case, and we should
treat such rare configurations with proper care.

\subsection{Estimation of differential rotation measure}

The  rotation measure and the excess position angle at
zero wavelength are computed from the best fit straight line between the
measured polarization position angle and the square of the
observed wavelength. The absolute rotation measure will have
the  uncertainty in the position angle by a multiple of $\pi$.
In the case of differential Faraday Rotation, this problem is,
by and large, absent and hence we expect that the differential rotation measure
is possibly a better indication of the magnetic field in the lens
than the absolute rotation measure. For typical Faraday rotation
in the case of nearby galaxies, the expected rotation of the
position angle can exceed $\pi$ at frequencies lower than
$\sim$ 5 GHz.
For the systems $0957+561$, $1938+666$ and $1830-211$, this
factor has been incorporated
and as such there is no ambiguity of $\pi$. The errors in the observed
position angles are available only for $0957+561$, and 
Patnaik (private communications) gives a value of 2$^\circ$ for the system
$0218+357$.  For systems like $1938+666$,
the errors are given only for one frequency and no error information
is available for other systems.
Consequently, the $\chi^2$ values shown in Table 1, with an assumed
error of 2$^\circ$ should
be taken with caution. However, the errors in the differential Faraday
Rotation Measure are not severely affected by this 
lack of error information in the data.
For instance, as 
\citet{king:97} have demonstrated
for the absolute Faraday Rotation for the system $1938+666$,
the fit and the uncertainty in the rotation measure are fairly 
good if we accept the data at all the frequencies with equal weight.
For differential Faraday rotation between the multiple images in 
gravitational lens systems, the excess position angle
at zero wavelength should be ideally zero, which provides an independent
check on the reliability of the fits given in Table 1.

\subsection{Faraday Rotation due to Spiral Galaxies}

In our sample we have two lens systems, B0218+357 and PKS1830--211,
where the lens is confirmed to be a spiral galaxy.
There is a large rotation measure $\sim$ 8000 rad m$^{-2}$ that is common to 
both the images A, B in the lens system B0218+357. 
The large intrinsic RM could be from absorbing clouds en route. On the other
hand, the relative fluxes of the two VLBI components change with frequency.
We cannot, therefore, rule out the possibility that part of the
change in the Faraday Rotation as function of frequency may be
caused by the presence of milliarcsecond substructures.
This is amply borne out by the detailed 8.4 GHz VLBA images of
\citet{bigg:99}. 
The polarization images of the core of both images exhibit
a difference of the order of 10 degrees in the polarization direction 
between the hot spots separated by a milliarcsecond. Consequently,
in the absence of detailed VLBA polarization maps, we may not be in a 
position to estimate the rotation measure at the source. However,   the 
difference in RM between images B and A, which is found to be
 980 $\pm$ 10 rad m$^{-2}$
 when the Faraday Rotation angle for four frequencies between 8.4 and 43 GHz
in 
\citet{patn:93}  
is used. This could conceivably be the
contribution of the lens galaxy. The result should be trustworthy
considering the error estimate, since for an uncertainty in the Position Angle of 
polarization vector of $\sim 2^\circ$, the $\chi^2$ is 0.65 for 2 degrees of 
freedom.
With the neutral hydrogen column density 
of $\sim 2\times 10^{21}$ cm$^{-2}$ and using the electron column density 
$N_e\sim 10^{21}$ cm$^{-2}$, we get from eq.~(3), the mean field 
magnetic component, along the sightline, in the lensing galaxy of 
order  $1\mu G$. The magnetic field could be even higher, provided the
relative RM  is not significantly overestimated.

\subsection{Differential Faraday Rotation in Elliptical Galaxies}

Both the lens systems Q0957+561 and B1422+231 have a giant elliptical galaxy
as the dominant lens. While the former is part of a galaxy-cluster,
the latter might be associated with a group of galaxies.
An analysis of the Faraday Rotation measurements in these and similar systems
could provide evidence for possible existence of coherent large--scale
magnetic fields in elliptical galaxies.

Two important problems need to be addressed: (a)effects of time delay
between the multiple images, which could become important for large time delays
 and (b)the Faraday Rotation introduced at the source.
Possibly the first lens system Q0957+561, with a time delay of $\sim$420 days
is a good example for examination of these issues.
The earlier measurements by
\citet{green:85}
for the system
Q0957+561 reported rotation 
measures of A and B images to be respectively -61 $\pm$ 3 rad m$^{-2}$ 
and -160 $\pm$ 3 rad m$^{-2}$. This gives difference in the RM between A 
and B images of $\sim$100 rad m$^{-2}$ which is at variance with the 
corresponding difference of $\sim$30 rad m$^{-2}$ measured by
\citet{patn:01}. 
Indeed, 
\citet{nair:93} had pointed out the importance of polarization 
measurements for time--delay estimations. Following the work of
\citet{bigg:99},
 the well established time variability of 
the polarization vector was used by Patnaik and Narasimha \citet{nar:01}
to numerically derive the time delay between the images in the system 0218+357,
thereby demonstrating the possible effects of time--delay on the
differential Faraday Rotation.
The discrepancy between 
\citet{green:85} and
\citet{patn:01} 
might be indicative of time--variability of the polarization vector
 in the system Q0957+561 over a time-scale of a decade.
This is a major problem while comparing polarization position angles at two
or more different epochs.

There is also a need to discuss contribution of the intervening medium 
to the Faraday Rotation.
\citet{green:85} 
attributed the difference between the 
rotation measures along the sightlines to the two images, of 
$\sim$ 100 rad m$^{-2}$ entirely to the lensing cD galaxy with intracluster 
medium making a negligible contribution.
This seems to be a reasonable deduction as also implied by our
best fit excess P.A. of 2 degrees.
\citet{perry:93} 
argue that because 
of the availability of separate rotation measures along two sightlines to 
images A, B, it should be possible to identify the contribution from the 
absorption line systems detected en route at redshifts $z_{abs}=1.39$ 
and $z_{abs}=1.12$. They speculate that the RM of -63 rad m$^{-2}$ that 
is common to both images A and B should be assigned to the absorption system 
located at $z_{abs} =1.39$ because of its large inferred electron column 
density. This implies $(N_e)_{19}<B_\parallel>_{\mu G}\simeq 130 cm^{-2}\mu G$. 
With the reported $N_e \approx 1.25 \times 10^{20} cm^{-2}$, 
the mean line-of-sight magnetic field in the lensing galaxy is inferred 
to be $< B_\parallel \approx \geq 10 \mu G$.
On the other hand, the
common Rotation Measure might merely be an artifact of the
substructure in the source. A systematic analysis of the Rotation Measure
along the multiple images of an extended feature like a jet, 
similar to what 
\citet{b26} 
did for a single image, will help
resolve this important issue.

Possibly the source structure in the lens system B1422+231
has similarities with B0218+357 at VLB scales, and, not surprisingly,
straightforward estimation of the absolute Faraday Rotation
 along individual images 
results in a large value of RM for both these systems.
The source in B1422+231 has smaller fraction of polarization and
hence, the change of polarization direction within the 
milliarcsecond scale structures is expected to affect 
the rotation measure estimates. Interestingly, the differential RM is 
coincidentally similar to the value arrived at for 0957+561, 
where the rotation is probably caused by a similar giant elliptical galaxy
at comparable redshift. The different ratio of flux between the images
in radio and optical is possibly an indication of extinction.

The nature of the lensing object  in the systems MG1131+0456 and
1938+666 has remained enigmatic for a long time, although the redshifts
have now become available
\citet{koch:00},\citet{t45}.
The main lens in both the systems appears to be a passively evolving giant
elliptical, but probably there are two clusters or rich groups of galaxies
in the field of MG1131+0456.
Still, we believe that Rotation Measure estimates for these
systems are important due to the presence of extended structures in the
background source. We expect Einstein Rings and giant arcs to provide 
valuable probes of the large--scale
magnetic field in the intervening object because, in principle, we can
trace the gradual variation of the position angle of the field vector
along the quasi--linear image structure. With better long term observations
of these systems, we should
be able to determine the length scale, signature and strength of magnetic field
in the lenses.

For the system MG1131+0456, the reported Faraday Rotation measurements
\citet{chen:93} are primarily for bright regions of the Einstein Ring.
It is conceivable that these features are, perhaps not identifiable with images of the same source-region and hence we cannot speculate on intrinsic
magnetic field of any absorbers near the source.
However, based on their detailed analysis (cf. Fig.6 \citet{chen:93}), 
we speculate that the difference in Rotation Measure between  spots R1 and R4,
of $\sim$ 1200 rad m$^{-2}$,
is an indication of the presence of a magnetic field in the lens galaxy.
This value is practically similar to what is estimated for B0218+357.
As emphasised earlier, Faraday Rotation measurements
for the system MG1131+0456 will be valuable even if the image identification 
may not be very robust.

For the 4-image system 1938+666,
the images C1 and C2 are highly magnified and
are at sub--arcsecond separation; so they are unlikely to be affected
by many of the other systematics discussed earlier.
They have a small differential rotation measure of 56 rad m$^{-2}$,
for the two images separated by approximately 5 kiloparsecs at the lens
and the excess position angle extrapolated to zero wavelength is negligible.
But they have a common Rotation Measure of
almost 500 rad/m$^2$ which is also seen in the other two images.

Thus, based on the differential Faraday Rotation between the
multiple images in 0957+561, 1938+666, MG1131+0456 and B1422+231,
there appears to be suggestive evidence that {\it giant elliptical galaxies may
also have ordered magnetic field with strength comparable to
that of spiral galaxies.}
The fit for differential Faraday Rotation is overall reasonable, 
but with lower fraction 
of polarization, it is certainly not as good as in the case of B0218+357.
In spite of the range of Faraday Rotation measure common to the images,
it is remarkable that for the four systems having small image separation,
the differential Faraday Rotation (introduced by the main lensing
galaxy alone) is of order 1000 rad m$^{-2}$ for systems with
vastly different properties.
This possibly suggests that by the redshift of 1, most of the magnetic fields
we see in galaxies (of the order of a few microgauss) which were presumably
generated in the pre-galactic epochs, might have become saturated.
It is tempting to surmise that perhaps already by redshift $\sim$1,
magnetic fields of strength of the order of $\mu$G are generated
over length scales of $\sim$10 kpc in the Universe.
However, we should emphasise again that we have put together data on polarized
images taken by various groups for different purposes and hence, the result
may not be as robust as we might hope.
Kronberg \citet{b24} emphasised the need
for co-ordinated VLBA/VLBI observations of extended multiply--imaged
systems to overcome many of the defects we have mentioned, and make
an attempt to get a reliable magnetic field profile at least in a few
lens systems.

It should, of course, be conceded  that the present method based on 
differential Faraday Rotation measure maps is not without its limitations. 
The primary reason for the inadequacy
of the data is that observations were not intended for the measurement
of the lens magnetic fields. However, considering the importance of
probing the large--scale cosmic magnetic fields at high redshift, 
we have used the
available data for deducing the existence of ordered magnetic field in external
galaxies. Clearly, a determination of the large--scale magnetic 
structures in the Universe at all scales is a fundamental problem in astronomy
and to address this question it is imperative to undertake coordinated 
multifrequency, multi--epoch VLBI/VLBA observations.
A selection of non-varying sources (e.g. knots) will help alleviate some of the
problems associated with time delays between the images.

\section{Case for ordered Magnetic Fields in Clusters?}

The existence of $\mu$G global fields on kpc scale, that are almost
aligned along the spiral structure 
has been observationally well established \citet{b1}.
It is evident from the foregoing discussions that probably the global
magnetic field we see in the nearby spirals is not very different from
that found in galaxies at redshift of 0.5 to 1.
We have attempted to demonstrate that there is evidence suggesting the existence
of magnetic fields of microgauss strength, coherent over
tens of kiloparsec scale even in giant elliptical galaxies.
Naturally, as the next step it is worthwhile searching for global scale
cluster magnetic fields. Galaxy--clusters
are the largest gravitationally bound systems and probably, the rich 
galaxy--clusters virialise after the formation of spiral galaxies.

Many independent observations over the past decade
have provided evidence for cluster magnetic field of the order of $\mu$G
\cite{b22, b9}.
The synchrotron emission from radio halos
associated with several galaxy clusters has now been detected. In particular,
the radio halos of the Coma and several other clusters have been extensively
studied recently to find that the halos have typically sizes $\sim$ Mpc
and are concentrated close to the centre of X-ray emission \citet{b19}.
Indeed, observations of the diffuse radio source
in Abell 85 showing enhanced X-ray emission was effectively used by
\citet{bagchi:98} 
to deduce a magnetic field of $\sim$1$\mu$G.
The estimates of magnetic fields in cluster halos,
based on minimum energy arguments 
\citet{b28} 
range from a fraction
of a microgauss to one microgauss. Thus, from radio observations of the 
Coma cluster \citet{beck:00},
a lower bound on the magnetic field of a tenth of a
microgauss in the halo region has been placed \citet{b37}.
A similar bound has also been obtained from the gamma-ray flux above 100 MeV
observed by EGRET \citet{b40}.
Remarkably, the magnetic field
strength derived by 
\citet{b20} 
using the minimal energy arguments also
yields similar estimates for the field.

It should be recognized that the results for intracluster {\it global} fields within the
galaxy--clusters are not so robust,
although there is reliable information
on the magnetic fields in member radio galaxies in the clusters
\citet{brand:05}.
An impressive study of a sample of sixteen
Abell clusters was undertaken by 
\citet{b12} 
to probe intracluster magnetic fields using radio and X-ray data.
From a statistical analysis of 
cluster sources situated in the hot cluster gas and of the controlled 
background sources located behind the cluster medium, they
found evidence for intracluster magnetic fields in the Abell clusters of 
order $\sim$ 5$\mu$G ($l_B/10$ kpc)$^{-1/2}$, $l_B$ 
being the coherence scale-length of the magnetic field.
This led them to conclude that
the high Faraday Rotation in embedded radio sources in galaxy--clusters
indeed originates from the foreground ICM
\citet{b9}, \citet{clark:04}, \citet{b23}.
However,
\citet{rudn:03}
argue the case in favour of source--local magnetic fields.
Greenfield, Roberts \& Burke(1985) had earlier pointed out the use of a
radio lobe associated with the lensed image 0957+561A to rule out a 
significant contribution to the Rotation Measure from the ICM of the 
lens cluster for this system.

It is clear that Faraday Rotation studies of both radio galaxies in clusters as well as 
background radio sources aligned with the cluster 
 can provide valuable probes for
 cluster magnetic fields.  Thus, in the core of the cooling 
flow cluster 3C295, a magnetic field of order 12$\mu$G has been estimated 
with  a coherence scale-length of about 5--10 kpc based on the patchiness of 
RM \citet{allen:01} somewhat similar to the value given
for 3C129 \citet{b7}.
 The multi--frequency observations of the radio galaxies embedded
in the X-ray cluster 3C129 reveal significant difference in the Faraday
RMs towards radio galaxies 3C129.1 located at the cluster centre
and 3C129 at its periphery, implying cluster magnetic field strength
of $\sim$ 6 $\mu$G out to a distance of $\sim$ 450 kpc.
There is observed a remarkable 
trend of the rotation measures in Hydra A which is positive to the north 
of the nucleus and negative to the south indicating a field strength 
$\sim$ 7$\mu$G with a coherence scale of $\sim$ 100 kpc \citet{tayl:93}.
The Faraday Rotation measurements of multiply--imaged
extended background sources will be valuable to probe this kind of structures.
The differences in RMs between various images should filter out contributions 
from the background source and from our Galaxy with the residual RM 
providing a value of $N_e<B_{11}>$ for a suitably averaged line-of-sight 
magnetic field component. An independent estimate of the electron column 
density from the thermal X-ray emission or from the measurements of 
Sunyaev--Zel'dovich  effect would provide an estimate of the magnetic field
strength in the cluster and the change in the field strength along the
images of an extended structure will, further pinpoint the coherence length
of this field \citet{b13}.

Ideally, we should attempt to search for radio arcs in well-studied Abell 
Clusters and measure the Faraday rotation along the arcs 
in order to deduce intracluster magnetic field strength and its coherence 
length. Assuming there is no active radio source in the lensing cluster 
contributing to the RM, we should enquire if we can deduce an intracluster 
magnetic field of strength $\sim$ 0.1 $\mu$G coherent over a scale-length 
exceeding 100 kpc.
Such an exploration will be facilitated if we should be able to locate 
magnetized sources such as polarized radio jets, starburst galaxies
and the associated synchrotron jets and 
even supernovae in distant galaxies which are
suitably aligned behind the foreground lensing clusters.
A multiply imaged polarized Gamma Ray Burst (GRB) source together with its 
afterglow in radio located behind a foreground cluster lens would also 
serve a useful purpose of indicating the strength of the intervening cluster 
magnetic field. 

It will be valuable to have radio maps in polarization at a few closely 
separated frequencies to estimate the RMs.
A candidate lens system like $B1359+154$, which  is a six--image 
configuration of a high redshift source 
$(z_{source}=3.235)$ having a flux of 66 mJy at 5 GHz frequency (with 
unknown lens redshift) could have the advantage of tracing
the large-scale magnetic field of the intervening lens. Due to the
six image nature of the system, the problem associated with
time delay between the images might be alleviated.
It will be useful to examine unconfirmed lens candidates, especially 
radio-weak systems with substantial image separations. 
These could serve as ideal clusters mass dark lenses for probing the 
intracluster magnetic fields. Amongst the six dozen or so lensed quasars 
discovered to date most have few arcsecond image separations which can be 
produced by galaxy-mass lenses. However, theoretical CDM models of structure 
formation predict large quasar image separations of several arcsec capable 
of being generated by dark matter aggregates. Indeed, 
\citet{b16}
have recently reported the discovery of a quadruply lensed quasar with a 
maximum image separation of 14.62 arcsec, an evident case of gravitational 
lensing by a dark matter dominated intervening object. There should be 
several such large-separation image configurations with hitherto undetected 
dark cluster mass objects acting as lenses.

A favourable set of observations for Faraday Rotation of lensed background
sources as well as embedded  radio sources within the cluster,
for a range of cluster redshifts, could provide valuable information
on the possible origin of magnetic fields in galaxy clusters:

\begin{enumerate}
\item{} Seed magnetic field produced during the protocluster formation
\citet{b4}. This
will be a global field spread over Mpc scale which is likely to be rather weak,
unless clusters grew by mergers.

\item{}Field produced by embedded radio sources:
The jet propagating into the intracluster medium will introduce an oriented
field, the strength of which will decrease away from the radio source.
\item{}Local fields due to anisotropic electron velocity distribution
\citet{okabe:03}:
Chandra observations of rich galaxy clusters show evidence
for sharp discontinuity in the density of X-ray emitting regions
as well as temperature over length scales of a few hundred kpc
in systems like Abell2142
where the electron anisotropic velocity could drive a Biermann current.
\end{enumerate}
A systematic observational study of  these three effects will be
valuable in understanding the substructures in galaxy clusters.

 \section{\bf Conclusions}
 
The multiply-imaged gravitational lens systems can be
effectively used to establish the presence of global ordered magnetic fields
in lensing galaxies and to estimate their average field strength. It
turns out that the difference in Position Angles between various
images is generally a reliable indicator of the existence of magnetic
fields in intervening lenses.
An advantage of multiple path RM measurements is that they are
potentially capable of sensing the direction as well as coherence length-scale
of the magnetic field. 
We have further argued that the contributions due to inhomogeneities in
our Galaxy need not be a major hurdle in estimating these
ordered magnetic fields, unless the depolarization effects between images
turns out to be substantial. The compact flat--spectrum sources will have
substructures in polarization which will naturally be subject to
differential magnification across the image, if the source is in the vicinity of
a caustic. 
Evidently, a non--varying polarized radio source would be ideal
for diferential Faraday Rotation measurements and equally to monitor
the lensed images for polarization changes in order to correct for the
time--delay.
The main conclusions of our study are the following:
\begin{enumerate}
\item{}There appears to be suggestive evidence for
the presence of coherent, large scale 
magnetic field in the lens systems we have examined, in particular, in giant
elliptical lens galaxies.

\item{}Substantial amount of Rotation Measure common to all the images is
observed in almost all the cases, which
probably  originates in the medium in the neighbourhood
of the source or may even be a result of not resolving the source
substructures.

\item{}In spite of a range of absolute Rotation Measures for the various systems
and along different images, the differential Rotation Measure appears
to be in the range of several hundred rad m$^{-2}$ for the 
elliptical galaxy lenses
and $\sim$ 1000 rad m$^{-2}$ for the spirals. 

\item{}The available sample of lens systems do not seem to indicate
any obvious evolution with redshift of the observed
Rotation Measure.
\end{enumerate}

\section*{Acknowledgements}
We are grateful to
Professor P. Kronberg for his encouragement and valuable comments
which have led to a considerable improvement in the manuscript.
We thank Professors J.P. Ostriker, A. Olinto, G. Memon  and K. Subramanian
for several useful discussions.
DN is grateful to Japan Society for the Promotion of Science for an
Invitational Research Fellowship. SMC is grateful to the DAE--BRNS
Senior Scientist scheme for a fellowship and to the Institute of
Astronomy, Cambridge for hospitality.


\label{lastpage}


\begin{thebibliography}{}
\bibitem[\protect\citeauthoryear{Beck et al}{2004}]{b1}
 Beck, R., 2004, ApSpS, 289, 293
\bibitem[\protect\citeauthoryear{Hogan}{1983}]{b2}
Hogan, C.H., 1983, Phys. Rev. Lett., 51, 1488
\bibitem[\protect\citeauthoryear{Turner \& Widrow}{1998}]{b3}
 Turner, M.S. and Widrow, L.M., 1988, Phys. Rev., D37, 2743
\bibitem[\protect\citeauthoryear{Subramanian et al}{1994}]{b4}
 Subramanian, K., Narasimha, D. and Chitre, S.M., 1994, MNRAS, 271, {\sc p}15
\bibitem[\protect\citeauthoryear{Kulsrud et al}{1997}]{b5}
 Kulsrud, R.M., Cen, R., Ostriker, J.P. and Ryu, D., 1997, ApJ, 480, 481
\bibitem[\protect\citeauthoryear{Kronberg \& Perry}{1982}]{b6}
 Kronberg, P.P. and Perry, J.J., 1982, ApJ, 263, 518
\bibitem[\protect\citeauthoryear{Taylor et al}{2001}]{b7}
 Taylor, G.B., Govoni, F., Allen, S.W. and Fabian, A.C., 2001, MNRAS, 326, 2
\bibitem[\protect\citeauthoryear{Taylor \& Perley}{1993}]{tayl:93}
 Taylor, G.B. and Perley, R.A., 1993, ApJ, 416, 554.
\bibitem[\protect\citeauthoryear{Blasi et al}{1999}]{blasi:99}
 Blasi,P, Burles, S, Olinto, A., 1999, ApJ Letters., 514, L79
\bibitem[\protect\citeauthoryear{Sargent et al}{1989}]{sarg:89}
Sargent, W.L.W. et al, 1989, ApJS, 69, 703
\bibitem[\protect\citeauthoryear{Walsh et al}{1979}]{walsh:79}
Walsh, D., Carswell, R.F. and Weymann, R.J., 1979, Nature,  279, 381
\bibitem[\protect\citeauthoryear{Patnaik et al}{1993}]{patn:93}
 Patnaik, A.R., Menten, K.H., Porcas, R.W., Kemball, A.J, 1993, 
MNRAS, 261, 435
\bibitem[\protect\citeauthoryear{Subrahmanyan et al}{1990}]{subh:90}
Subrahmanyan, R., Narasimha, D., Rao, A.P., Swarup, G., 1990, MNRAS, 246, 263
\bibitem[\protect\citeauthoryear{Greenfield et al}{1985}]{green:85}
 Greenfield, P.E., Roberts, D.H. and Burke, B.F., 1985, ApJ, 293, 370
\bibitem[\protect\citeauthoryear{Patnaik et al}{1999}]{patn:99}
 Patnaik, A.R. {\it et al.}, 1999, MNRAS,  307, 1{\sc p}
\bibitem[\protect\citeauthoryear{Patnaik et al}{2001}]{patn:01}
 Patnaik, A.R., Menten, K.H., Porcas, R.W. and Kemball, A.J., 2001, 
in Gravitational Lensing: Recent progress and future goals. ASP, proceedings,
eds. T.G. Brainerd, C.S. Kochaneck, {\sc p}99 
\bibitem[\protect\citeauthoryear{King et al}{1997}]{king:97}
King, L.J., et al, 1997, MNRAS, 289, 450
\bibitem[\protect\citeauthoryear{Chen \& Hewitt}{1993}]{chen:93}
 Chen, G.H. and Hewitt, J.N., 1993, AJ,  106, 1719
\bibitem[\protect\citeauthoryear{Patnaik \& Narasimha}{2001}]{nar:01}
 Patnaik, A.R. Narasimha, D., 2001, MNRAS, 326, 1403
\bibitem[\protect\citeauthoryear{Patnaik}{1999}]{p30}
 Patnaik, A.R., 1999 (private communication)
\bibitem[\protect\citeauthoryear{Biggs et al}{2003}]{bigg:03}
 Biggs, A.D.{\it et al}, 2003, MNRAS, {\bf 338}, 599.
\bibitem[\protect\citeauthoryear{Rand \& Kulkarni}{1989}]{rand:89}
 Rand, R.J., Kulkarni, S.R., 1989, ApJ, 343, 760
\bibitem[\protect\citeauthoryear{Sofue et al}{1986}]{b39}
Sofue, Y., Fujimoto, M., Wielebinski, R., 1986, ARA \& A, 24, 459
\bibitem[\protect\citeauthoryear{Nair et al}{1993}]{nair:93}
Nair, S., Narasimha, D., Rao, A.P., 1993, ApJ, 407, 46.
\bibitem[\protect\citeauthoryear{Biggs et al}{1999}]{bigg:99}
Biggs, A.D. {\it et al}, 1999, MNRAS, 304, 349
\bibitem[\protect\citeauthoryear{Perry et al}{1993}]{perry:93}
 Perry, J., Watson, A.M. and Kronberg, P.P., 1993, ApJ, 406, 407
\bibitem[\protect\citeauthoryear{Kronberg et al}{1992}]{b26}
 Kronberg, P.P., Perry, J.J., Zukowski, E.L.H., 1992, ApJ, 387, 528
\bibitem[\protect\citeauthoryear{Kochanek et al}{2000}]{koch:00}
Kochanek, C.S. et al, 2000, ApJ, 535, 692
\bibitem[\protect\citeauthoryear{Tonry \& Kochanek}{2000}]{t45}
 Tonry, J.L., Kochanek, C.S., 2000, AJ, 119, 1078.
\bibitem[\protect\citeauthoryear{Kronberg}{2004}]{b24}
 Kronberg, P.P., 2004 (private communication)
\bibitem[\protect\citeauthoryear{Kronberg}{1994}]{b22}
 Kronberg, P.P., 1994, Rep. Prog. Phys., 57, 325
\bibitem[\protect\citeauthoryear{Carilli \& Taylor}{2002}]{b9}
 Carilli, C.L. and Taylor, G.B., 2002, ARA\&A, 40, 319
\bibitem[\protect\citeauthoryear{Kemper \& Sarazin}{2001}]{b19}
Kemper, J.C. and Sarazin, C.L., 2001, ApJ, 548, 639
\bibitem[\protect\citeauthoryear{Bagchi et al}{1998}]{bagchi:98}
Bagchi, J. et al, 1998, MNRAS, 296, L23.
\bibitem[\protect\citeauthoryear{Miley}{1980}]{b28}
 Miley, G., 1980, A \& A, 18, 165
\bibitem[\protect\citeauthoryear{Beck}{2000}]{beck:00} 
Beck R., 2000, Phil. Trans. Roy. Soc. Lond., A358, 777
\bibitem[\protect\citeauthoryear{Raphaeli et al}{1994}]{b37}
 Raphaeli, Y., Ulmer,M. and Gruber, D., 1994, ApJ, 429, 554
\bibitem[\protect\citeauthoryear{Sreekumar et al}{1996}]{b40}
Sreekumar, et al, 1996, ApJ, 464, 628
\bibitem[\protect\citeauthoryear{Kim et al}{1990}]{b20}
 Kim, K.-T., Kronberg, P.P., Decodney, P.E. and Landecker, T.L., 1990, ApJ,
355, 29
\bibitem[\protect\citeauthoryear{Brandenberg \& Subramanian}{2005}]{brand:05}
 Brandenburg, A. and Subramanian, K., 2005, Phys. Rep., 417, 1.
\bibitem[\protect\citeauthoryear{Clarke et al}{2001}]{b12}
 Clarke, T.E., Kronberg, P.P. and Bohringer, H., 2001, ApJL, 547, L111
\bibitem[\protect\citeauthoryear{Clarke}{2004}]{clark:04}
 Clarke, T.E., 2004, J. Korean Astron. Soc., 37, 337.
\bibitem[\protect\citeauthoryear{Kronberg}{2004}]{b23}
 Kronberg, P.P., 2004, J. Korean Astron. Soc., 37, 501.
\bibitem[\protect\citeauthoryear{Rudnick \& Blundell}{2003}]{rudn:03}
Rudnick, L. and Blundell, K.M., 2003, ApJ, 588, 143
\bibitem[\protect\citeauthoryear{Allen et al}{2001}]{allen:01}
 Allen, S.W et al., 2001, MNRAS, 324, 842
\bibitem[\protect\citeauthoryear{Ensslin \& Vogt}{2003}]{b13}
 Ensslin, T., Vogt, C., 2003, A\& A, 401, 835
\bibitem[\protect\citeauthoryear{Inada et al}{2003}]{b16}
Inada, N. et al, 2003, Nature, 426, 810.
\bibitem[\protect\citeauthoryear{Okabe \& Hattori}{2003}]{okabe:03}
 Okabe, N., Hattori, M., 2003, ApJ, 599, 964

\end{thebibliography}
\end{document}